\documentclass[aps,pre,twocolumn,nopacs,amssymb]{revtex4}

\usepackage{amsmath}

\usepackage{graphicx}
\usepackage{dcolumn}
\usepackage{bm}
\usepackage{color}

\usepackage{amsmath}
\usepackage{amssymb}

\def\be{\begin{equation}}
\def\en{\end{equation}}
\def\bea{\begin{eqnarray}}
\def\ena{\end{eqnarray}}
\def\bec{\begin{equation}\begin{array}{rcl}}
\def\p{\partial}

\def\gs{\gtrsim}
\def\ls{\lesssim}

\def\ab{{ij}}

\newcommand{\av}[1]{\langle{#1}\rangle}
\def\ab{\alpha\beta}

\newcommand{\bi}[1]{\mbox{\boldmath$#1$}}

\begin{document}
\title{Ions and dipoles in electric field: Nonlinear polarization  and 
field-dependent chemical reaction}  
\author{Akira Onuki\footnote{e-mail: onuki@scphys.kyoto-u.ac.jp 
(corresponding author)}}

\affiliation{
 Department of Physics, Kyoto University, Kyoto 606-8502, Japan 
}


\date{\today}

\begin{abstract} 
We investigate  electric-field effects in dilute 
 electrolytes with nonlinear polarization. 
As a first example of such systems, we  
  add   a dipolar component with a relatively large 
dipole moment $\mu_0$ to  an aqueous electrolyte.  
As a  second example,  the solvent 
itself exhibits nonlinear polarization near charged objects.  
 For such systems, we present a Ginzburg-Landau 
free energy and   introduce  field-dependent 
chemical potentials,  entropy density, and  stress tensor,  
which satisfy  general  thermodynamic relations. 
In the first example, 
 the dipoles accumulate in  high-field regions, 
as predicted by  Abrashikin {\it et al}.$[$Phys.Rev.Lett. 
{\bf 99}, 077801 (2007)$]$.  Finally, 
we consider the case, where    Bjerrum ion pairs 
form a  dipolar  component with  nonlinear polarization. 
The   Bjerrum dipoles accumulate in  high-field regions,   
 while   field-induced  dissociation was predicted by Onsager 
 $[$J. Chem. Phys.{\bf 2}, 599 (1934)$]$.
 We  present an  expression for 
the field-dependent association constant 
$K(E)$, which  depends on the field strength 
 nonmonotonically. 
 \end{abstract}


\maketitle


\section{Introduction} 

The electrostatic interactions among 
electric charges and dipoles in a solvent 
are of central importance 
in various situations in soft matter physics \cite{Russel,Stokes,Is}. 
In this paper, we consider 
dilute electrolytes composed 
of  a waterlike liquid solvent, cations, anions, 
 and a dipolar component with a  dipole moment $\mu_0$. 
   Andelman, Orland, and their coworkers (AO) 
\cite{An1,An2,An3,An4} 
proposed a dipolar Poisson-Boltzmann (dPB) equation, 
where the dipoles  can   respond to electric field nonlinearly  
 and their polarization density ${\bi p}_d$ 
yields the effective charge density $-4\pi\nabla\cdot{\bi p}_d$. 
  In their papers the dipolar component is   
the solvent itself. Theoretically,    it  can be different 
from the solvent with some changes of the equations.  
The dPB equation  is a generalization of the classical 
Poisson-Boltzmann equation 
and is convenient to investigate the nonlinear polarization 
around charged objects.  
 In particular, AO group  concluded  that the dipole density 
is increased by the factor 
$\sinh h/h$ with $h= \mu_0 E/k_BT$ \cite{An1}, 
where $E=|{\bi E}|$ is the field strength, $T$ is the temperature,  and $k_B$ 
is the Boltzmann constant. In this paper we show the 
following. The  dipole accumulation in high-field regions 
occurs if the dipolar component 
is a dilute solute in a solvent. On the other hand,  
 a nearly incompressible polar solvent is  hardly enriched  
 in high-field regions.

The physics of ion-dipole systems is  
even more intriguing if 
associated ion pairs, Bjerrum 
dipoles \cite{Bje,Ma,Fuoss,Roij,Dill,Vegt}, 
are treated to form a dipolar component. In electrolytes, 
 association of free ions 
and dissociation of bounded ion pairs balance  on the average 
 in equilibrium, while strong acids in   full dissociation  
have   long  been studied since the seminal 
work by  Debye-H\"uckel \cite{Russel,Is,Debye,McQ,Stokes}. 
Although not well known,   Onsager\cite{Onsager34} 
predicted that   applied electric field   enhances  
breakage of the ion bonding to increase  
  dissociation. He intended  to explain   
conductance increases in  weak 
electrolytes under  electric field  (the Wien effect) 
 \cite{Wien,Eck,Kaiser,Korea}.  In his theory, 
the  effect is significant 
for $E\gs 2k_BT/e\ell_B(= 0.08$V$/$nm  
for ambient water),   where  $e$ is the elementary charge 
and $\ell_B$ is  the Bjerrum length.
Therefore, in high-field regions,  some fraction of the 
Bjerrum   dipoles  dissociate in the Onsager  theory, 
while dipoles of a dilute dipolar component  accumulate 
without chemical reactions in the AO theory.
  
In treating the Coulombic and dipolar interactions, 
AO group used a   Hubbard-Stratonovich transformation 
of  the grand-canonical partition function. 
In this paper, we propose a  Ginzburg-Landau  
free energy  of  electrolytes, 
where the polarization can be  nonlinear  in  inhomogeneous  
 electric field. With this   approach we can 
 draw   unambiguous conclutions  systematically.    
Because the problem is   complex, 
we divide this paper into three parts:  
(i)  A  dipolar component is added 
as a dilute solute,  
(ii)  the solvent exhibits nonlinear polarization in Sec.III, and  
(iii)   Bjerrum ion pairs  form a dipolar component  in Sec.IV. 
There is no chemical reaction  in the first two parts.

In our problem, we need to develop a thermodynamic  theory 
of inhomogeneous electrolytes. In their book, 
 Landau and Lifshits  \cite{Landau-e} 
  examined  thermodynamics  of dielectric fluids  
 in electric field, providing  explanations of 
the electrostriction effect \cite{Hakim,Stell}  
and the Maxwell stress tensor. 
It shows   that the stress tensor  
is not diagonal and the  scalar pressure is not 
well-defined in electric field in such fluids. 
In this paper, we use the Maxwell stress tensor for electrolytes 
and  introduce  field-dependent  
chemical potentials $\mu_i$  and 
 entropy density $S$.  We find that these variables 
  satisfy  general  thermodynamic relations. 
For  Bjerrum dipoles,  we  determine   the field-dependent    
association constant $K(E)$ using the field-dependent 
$\mu_i$.  In the literature 
\cite{Bje,Ma,Fuoss,Roij,Dill,Vegt,Kirk,Landau-s}, 
the zero-field association constant $K(0)$ has been discussed.
We can then  
examine   the  distributions of ions and dipoles near charged 
surfaces, incorporating the AO  and 
 Onsager theories.

 The organization of this paper is as follows. In Sec.II, 
 we will present a theromodynamics theory 
   of dilute electrolytes consisting of four-components in 
electric field. 
In Sec.III, we will examine  dilute electrolytes consisting of 
a polar solvent and a strong acid, 
where the solvent exhibits nonlinear polarization.  
In Sec.IV, we will investigate association and 
dissociation in electrolytes in electric field, where 
Bjerrum dipoles  constitute the fourth component.   
In Appendix A, we will derive  the electric free energy 
density using the Onsager theory of 
rodlike molecules \cite{Onsager49}.

\section{Thermodynamics of  dilute  electrolytes in electric field}

In this section, we treat a four component system  
composed of a single-component polar solvent (1), 
cations with charge $Ze$ (2), 
anions with charge $-Ze$ (3), and a dipolar 
component with  a molecular dipole moment $\mu_0$ (4). 
Here,  $Z=1$ for monovalent salts (NaCl) and  
 $Z=2$ for divalent salts (MgSO$_4$). 
We assume that the solvent dipole moment is  considerably 
 smaller  than $\mu_0$ and the ions have no  dipole moment, 
so the solvent 
exhibits linear polarization.  As a typical solvent, 
we can suppose  ambient liquid water 
at $T = 300$ K and  $p = 1$ atm.  
In this paper, we use the  cgs units.

\subsection{Polarization of dipolar component}

In our continuum theory, 
the densities $n_i({\bi r})$ of the four components 
are coarse-grained ones   with 
 spatial scales longer than the Bjerrum length  $\ell_B  
= e^2/\epsilon_{\rm w} k_BT$, where  $\epsilon_{\rm w}$ 
is the  pure-solvent dielectric constant. 
The electric potential $\Phi$ 
and the  electric field ${\bi E}= -\nabla\Phi$ 
  also vary smoothly. 
To apply electric field, we suppose 
  parallel charged   plates   
separated by a  distance   $L$, within which  
 an electrolyte is inserted  in the region  $0<z<L$. 
 The lateral dimensions of the container 
  much exceed  $L$ 
and  the edge effect is negligible.

The charge density $\rho$ and the polarization  
${\bi p}_d$ of  the dipolar component 
are written as 
\be 
\rho=Ze(n_2-n_3),~~~ {\bi p}_d= n_4{\bi \mu}_d, 
\en 
where  ${\bi \mu}_d({\bi r})$ is the   
 average  dipole moment vector  of the dipolar 
component per particle, as will be defined in  Eq.(7) below. 
The  polarization of the solvent molecules  
is expressed in the linear response form, 
\be 
{\bi p}_{\rm w}= (4\pi)^{-1}
(\epsilon-1){\bi E}.
\en   
 We then introduce the electric induction, 
\be 
 {\bi D}={\bi E}+4\pi ({\bi p}_{\rm w}+{\bi p}_d)
=\epsilon{\bi E} +4\pi {\bi p}_d, 
\en 
which is related to $\rho$ by  the Poisson equation, 
\be 
 \nabla\cdot{\bi D}=\nabla\cdot(\epsilon{\bi E} +4\pi {\bi p}_d)
=4\pi\rho. 
\en 

As shown  in  Eqs.(2) and (3),  $\epsilon$ is  the  linear 
dielectric constant  of  the solvent 
molecules in the static  limit. 
In dielectric experiments of   1:1 small salts 
 in  liquid water    
\cite{Co,Wei,Bu}, $\epsilon$ decreased 
as $\Delta\epsilon= 
\epsilon-  \epsilon_{{\rm w}}\cong -A_1 c $  
for  the salt concentration   $c$  below  1M$/\ell$,  
where     $A_1$ is  a  constant. 
Here,   $\Delta\epsilon$  
 ranged       from $-8$ to $-20$  at $c\sim 1$M$/\ell$ 
depending on the salt species. 
This large decrease  arises from the  formation of 
the hydration   shells 
of  water   around  ions. That is, 
 the water molecules inside the shells  do not freely rotate 
 due to the ion-dipole 
interaction and their contribution to 
  $\epsilon$ is largely suppressed. 
For small $n_i$ ($i\ge 2$) we thus express $\epsilon$ as  
\be 
\epsilon= \epsilon_{{\rm w}}(n_1,T) 
+ {\sum}_i' \epsilon_i^{{\rm w}}(n_1, T) n_i +\cdots,  
\en 
where    the  coefficients 
$\epsilon_i^{{\rm w}}$ 
are independent of $n_i$ $(i\ge 2$). 
Hereafter, ${\sum}_i'$ denotes summation 
over the solute species. Since $n_2=n_3$ on the average, 
the degree of hydration 
is  indicated by  the dimensionless coefficient,  
\be 
{\hat{\gamma}}= -n_1(\epsilon_2^{\rm w}+\epsilon_3^{\rm w})/2\epsilon_{\rm w},
\en  
which  is  positive   in the range $2.5\ls {\hat\gamma}\ls 6$ 
(even divided by $2\epsilon_{\rm w}=160$). 
Thus,  $\Delta\epsilon/\epsilon_{{\rm w}}
\cong -\hat{\gamma}(n_2+n_3)/n_1$ for 
 1:1 salts  with  $c\ls 1$M$/\ell$ in water. 
For larger $c$, 
Buchner {\it et al.} \cite{Bu} 
obtained 
 $\Delta\epsilon\cong  -A_1c+ A_2 c^{3/2}$ 
with $A_1$ and $A_2$ being positive constants. 
For NaCl they  argued that  
the degradation of $\epsilon$   is  mostly 
due to  Na cations.
In Eq.(5)  we  add the term $\epsilon_4^{\rm w}n_4$ 
 in case    the dipolar particles  
 affect the solvent polarization.  

The  dipole moment vector 
${\bi \mu}_d$ in Eq.(1) 
is the statistical average defined by  
\be 
{\bi \mu}_d= \mu_0 \av{{\bi n}_d}=\frac{\mu_0}{4\pi}  
\int \hspace{-1mm}d\Omega P({\bi n}_d, {\bi r})  {\bi n}_d ,
\en
where ${\bi n}_d$ is the microscopic unit vector  along 
the dipole direction,  $\int \hspace{-0.51mm}d\Omega$ 
is the angle integration 
of  ${\bi n}_d$, and  
$P({\bi n}_d, {\bi r}) $ is the angle distribution function 
at position $\bi r$ with $\int d\Omega P({\bi n}_d, {\bi r}) =
4\pi$. In Appendix A, we shall see that the equilibrium form 
of $P({\bi n}_d, {\bi r}) $ in the dilute reime is 
written in terms of the local electric field ${\bi E}({\bi r})$ as 
\be 
P({\bi n}_d, {\bi r}) =  {\cal N}_{\rm d}^{-1}
 \exp[\mu_0 {\bi n}_d\cdot{\bi E}({\bi r})/k_BT] ,
\en 
Here,  ${\cal N}_{\rm d}$ is the normalization factor,   
\be 
{\cal N}_{\rm d}=\frac{1}{4\pi} \int \hspace{-1mm}d{\Omega} 
 \exp[\mu_0 {\bi n}_d\cdot{\bi E}/k_BT]=\frac{ \sinh h}{h}  .
\en 
which depends on  the dimensionless field strength,  
\be 
h= \mu_0 E/k_BT ,
\en
with  $E=|{\bi E}|$. Here, we 
 introduce  $W(h)= \ln {\cal N}_{\rm d}$ and $W'(h)= 
dW(h)/dh$ expressed as 
\be 
W=\ln (\sinh h/h),~~~ W'= {\cal L}(h)=  \coth(h)- 1/h, 
\en  
where  ${\cal L}(h) $  is the 
Langevin function with  ${\cal L}\cong h/3$ for $h\ll 1$ 
and ${\cal L}\cong  1$ for $h\gg 1$.  
Then,   Eqs.(7)-(9) give    
\be 
{\bi p}_d=n_4 \mu_0{\cal L}(h) E^{-1}{{\bi E}}
 ={\chi}_d {\bi E}  .
\en 
Here,   ${\chi}_d(h)= n_4 \mu_0 {\cal L}(h) / E$ 
is the nonlinear  susceptibility 
of the dipolar component. 
Setting ${\bi D}= \epsilon_{\rm eff}{\bi E}$, 
we obtain    the total nonlinear  dielectric constant, 
\be 
\epsilon_{\rm eff}= \epsilon+4\pi \chi_d
=  \epsilon+4\pi n_4 \mu_0 {\cal L}(h) / E. 
\en 
As $h\to 0$ it follows the total linear dielectric constant,  
\be   
\epsilon_{\rm eff}^0={\lim}_{h\to 0}\epsilon_{\rm eff}
= \epsilon+  4\pi(\mu_0^2/3k_BT)n_4.
\en    
For $h\gg 1$, the dipolar particles tend to fully 
align along $\bi E$ with 
${\chi}_d \cong  n_4\mu_0/E $. 

AO group \cite{An2,An4} 
calculated $\epsilon_2^{\rm w}$ and $\epsilon_3^{\rm w}$  
using   their  dPB  
equation around ions, where 
the solvent (water) itself is the dipolar component.   
In our scheme, we  assume their presence 
with large sizes from the beginning.   
 Frydel \cite{Frydel} constructed the dPB equation for  ions 
with polarizabilities $\alpha_i$.  
The ionic  polarization  can be 
accounted for if   we  change $\epsilon_i^{\rm w} $ 
to $\epsilon_i^{\rm w} +4\pi\alpha_i$ in Eq.(5).  
However,   $|\epsilon_2^{\rm w}+\epsilon_3^{\rm w}|$ 
is  much larger than $4\pi\alpha_i$ for small ions\cite{Madden}.

\subsection{Free energy functional and chemical potentials}

We set up  the  free energy functional  ${\cal F}=\int d{\bi r}f$, 
where the integration is within the cell confining the 
electrolyte solution. We neglect   the van der Waals  fluid-solid 
interaction and the image potential, 
which can be  important  near the surfaces, however. 
For    $n_i\ll n_1$ ($i\ge 2$), the Helmholtz 
free energy density $f$ is 
written as \cite{Onukipolar,anta,Ben,Oka1,Oka2}  
\be
 f  =f_{\rm w} + k_B T{\sum}'_{i} 
 n_i[\ln ( n_i\lambda_i^3) +\nu_i-1] + f_e ,
\en 
where $ f_{\rm w}( n_{1},T )$ is 
 the    pure-solvent free energy density  without applied field. 
The first two terms constitute the non-electric part.
For each $i (\ge 2)$, 
$\lambda_i (\propto T^{-1/2})$ is   the thermal 
de Broglie length and    $\nu_i(n_1,T)$  consists of 
 ideal-gas and  interaction parts. 
The former is written as  \cite{Landau-s},      
\be 
  \nu_i^0(T)=  -(k_i-3/2)\ln (T/\Theta_i^0) , 
\en 
where  $2k_i-3$ is   the  degree of  vibration-rotation freedom 
and    $\Theta_i^0$ is  a   constant temperature.  
The difference     $\Delta\nu_i= \nu_i-\nu_i^0$ 
is determined by   the solute-solvent interactions at infinite dilution.  
Although not written in Eq.(15), 
 the Debye-H\"uckel free energy 
density $f_{\rm DH}=-k_BT \kappa^3/12\pi$ \cite{Debye}  and 
 the short-range ion-ion interaction terms 
are important to explain experimental data for not very 
dilute electrolytes \cite{McQ,Stokes,Oka2},   
where $\kappa=Z[4\pi(n_2+n_3)\ell_B]^{1/2}$ 
is  the Debye wave number.

As will be shown  in Appendix A, we use Onsager's theory of 
rodlike molecules  \cite{Onsager49} to obtain  
 the electric free energy density $f_e$. 
Neglecting the short-range dipole-dipole 
interactions, we obtain    
\bea 
&&\hspace{-1cm}  f_e= \epsilon E^2/8\pi 
+ {\bi p}_d\cdot{\bi E}-  k_BTn_4 W\nonumber\\
&&\hspace{-6mm} =  \epsilon E^2/8\pi 
+   k_BTn_4( h W' -  W). 
\ena  
where   $W(h)$ and $W'(h)$ are defined in Eq.(11).  
 For $h\ll 1$, 
we have $W\cong h^2/6$ and $hW'\cong h^2/3$, so 
 $f_e$ tends to the linear response form,   
\be 
f_e^0=\epsilon_{\rm eff}^0 E^2/8\pi =
 (\epsilon+4\pi n_4\mu_0^2/3k_BT)E^2/8\pi.
\en  


Next, we  slightly change $n_i$ and  $T$ by $\delta n_i$ and $\delta T$. 
Then, from $\delta (\epsilon E^2)=2{\bi E}\cdot\delta{\bi D}  
-8\pi{\bi E}\cdot\delta{\bi p}_d -E^2\delta\epsilon$ 
and ${\bi p}_d\cdot\delta{\bi E}= n_4\mu_0 W'\delta E$, 
  $f_e$  is changed by    
\bea 
&&\hspace{-1cm}\delta f_e= 
{\bi E}\cdot \delta{\bi D}/4\pi-E^2\delta \epsilon/8\pi 
 \nonumber\\
&& -(k_BTW) \delta  n_4 
+ k_B n_4(hW'-W) \delta T.
\ena 
Here, integration of   the first term within the cell becomes 
\be 
\frac{1}{4\pi} \int d{\bi r} {\bi E}\cdot \delta{\bi D}
= \int d{\bi r} \Phi\delta\rho -
(\Phi_0 \delta Q_0+ \Phi_L\delta Q_L),
\en 
where $(\Phi_0,Q_0)$ and $(\Phi_L,Q_L)$ 
are the surface potential and the surface charge  
at $z=0$ and $z=L$, respectively.  
If we assume the overall charge neutrality, we require   
\be 
\int d{\bi r} \rho = Q_0+ Q_L,  
\en  
under which we can replace  $\Phi$ 
by $ \Phi-C_0$ with $C_0$ being an arbitrary constant. 
The  surface term in Eq.(20)  vanishes
 at fixed surface charges $(\delta Q_0=\delta Q_L=0$).

We  define the chemical potentials  $\mu_i$ by 
the functional derivatives 
$\mu_i=\delta{\cal F}/\delta 
 n_i $ at fixed $T$ and $n_j$ with $j\neq  i$. 
Then, Eqs.(15), (19),  and (20) give      
\bea 
&&\hspace{-8mm}
\mu_{1} /k_BT = \mu_{\rm w} /k_BT   
+ {\sum}'_{i} \nu'_in_i  - \zeta_1 ,\\
&&\hspace{-8mm}
 {\mu}_{2}/k_BT=   
  \ln (n_2\lambda_2^3)+\nu_2 + U-\zeta_2 ,\\
&&\hspace{-8mm}
 {\mu}_{3}/k_BT=   
  \ln (n_3\lambda_3^3)+\nu_3 -U-\zeta_3 ,\\
&&\hspace{-8mm}
\mu_4/k_BT =\ln (n_4\lambda_4^3)+\nu_4- W -\zeta_4,   
\ena 
where  $ \mu_{\rm w} (n_1,T)= \p f_{\rm w}/\p n_1$ 
is the chemical potential of pure solvent. For each $i \ge 1$  we define 
\bea 
&&\zeta_i= \epsilon_i'  E^2/8\pi k_BT,\\  
&&U= Ze\Phi/k_BT.  
\ena 
We write the density derivatives of  $\nu_i$ and $\epsilon$ at 
fixed $T$  as 
\bea
&&\nu'_i=\p \nu_i/\p n_1 ~(i\ge 2), \\
&& \epsilon_i'=\p\epsilon/\p n_i~ (i\ge 1),
\ena 
so $\epsilon_1'\cong 
\p\epsilon_{\rm w}/\p n_1$ 
and $\epsilon_i' \cong 
\epsilon_i^{\rm w}$ for  dilute solutes. 
In  Eq.(22),  the last term $-\zeta_1$ 
gives rise  to the electrostriction 
 (see Eq.(50)) \cite{Landau-e,Stell,Hakim,Kirk}. 
In Eq.(23) and (24), the  terms $-\zeta_i$ serve to 
increase $\mu_i$  in $\bi E$ for   
$\epsilon_i^{{\rm w}}<0$  (see below Eq.(5)).
In Eqs.(23)-(25),  $\nu_i $  
can be treated to be homogeneous constants in one-phase states. 
However,  from their strong dependence on $n_i$, 
 they exhibit discontinuities  across interfaces in  two-phase states, 
giving rise to  solute density differences in the two phases.

We also  define the entropy density as  the functional derivative 
$S= - \delta {\cal F}/\delta T$ at fixed  
densities  $\{ n\}$ and fixed surface charges. 
The integral of ${\bi E}\cdot\delta{\bi D}/4\pi$ 
does not contribute to $S $ from Eq.(20) so that  
\bea 
&&\hspace{-11mm}  S =S_{\rm w} 
 -k_B {\sum}'_{i} n_i\big[\ln (n_i\lambda_i^3)-\frac{5}{2}  
+ \nabla_T(T  \nu_i)\big]\nonumber\\
&&+\frac{1}{8\pi}E^2\nabla_T{\epsilon} -  k_B n_4(hW'-W),
\ena
where   $S_{\rm w}(n_1,T)= -\p f_{\rm w}/\p T$ 
is   the   pure-solvent part and the last  term is 
 the entropy density of  dipole  orientation (see Appendix A). 
  Here,  
$\nabla_T(\cdots)= \p(\cdots)/\p T$ denotes 
 the temperature derivative 
at fixed densities $\{ n\}$.

\subsection{Free energy variations }

As in Eq.(19) we consider  small changes 
in $n_i$ and $T$. Then, the free energy density $f$ 
in Eq.(15) is changed by   
\be 
\delta f=  {\sum}_i \mu_i\delta n_i -S\delta T-
\nabla\cdot(\Phi \delta{\bi D})/4\pi, 
\en 
where we use the field-dependent $\mu_i$  in Eqs.(22)-(25) 
and  $S$ in Eq.(30). 
If the cell volume $V$ is fixed, 
   the free energy 
increment $\delta{\cal F}=\int \hspace{-1mm}d{\bi r}\delta f$ 
is expressed  as 
\be 
\delta{\cal F}
= \int \hspace{-1mm}d{\bi r}\Big[{\sum}_i \mu_i\delta n_i -S\delta T\Big]
- \Phi_0 \delta Q_0- \Phi_L\delta Q_L,
\en 
Thus,  ${\cal F}$  is used   at fixed surface charges $\delta Q_0=\delta Q_L=0$. 

If  the potential difference 
$\Delta\Phi= \Phi_0-\Phi_L$ is fixed, we  perform 
the Legendre transformation \cite{Landau-e},  
\be 
\tilde{\cal F}= 
{\cal F}+ \Phi_0 Q_0+\Phi_L Q_L= {\cal F}+
 \int d{\bi r}\big[\Phi\rho -\frac{1}{4\pi}
{\bi E}\cdot{\bi D}\big].   
\en  
Setting $\tilde{\cal F}= \int d{\bi r}{\tilde f}$, 
we can define the electric free energy density   ${\tilde f}_e$ by
   \bea 
&&\hspace{-1cm}
{\tilde f}_e=f_e+\Phi\rho -{{\bi E}\cdot{\bi D}}/{4\pi}\nonumber\\
&&\hspace{-6mm}= \Phi\rho -{\epsilon E^2}/{8\pi} -  k_BTn_4 W.   
\ena 
The incremental changes in $\tilde f$ 
and $\tilde{\cal F}$  are given by  
\bea 
&& 
\delta {\tilde f}=  {\sum}_i \mu_i\delta n_i -S\delta T+ 
\nabla\cdot(\delta\Phi {\bi D})/4\pi, \\
&&\hspace{-1cm} 
\delta\tilde{\cal F}
= \int \hspace{-1mm}d{\bi r}\Big[{\sum}_i \mu_i\delta n_i -S\delta T\Big]
+ Q_0\delta\Phi_0 + Q_L\delta\Phi_L.
\ena 
For $h\ll 1$,  ${\tilde f}_e$  tends to 
$\Phi\rho -f_e^0$ with  $f_e^0$ 
being  given by Eq.(18),   as it should be the case.

Previously, Ben-Yaakov {\it et al.} \cite{Ben} used  
the electrostatic free energy density  composed of the 
first two terms in the second line of Eq.(34), while 
we set it equal to $\epsilon E^2/8\pi$, the first term in 
Eq.(17) \cite{Onukipolar,anta,Oka2}. 
In these papers,   the polarization 
was assumed to be linear.

\subsection{Maxwell stress tensor}

The solution  stress tensor  $\Pi_{\ab}$ 
consists of two parts, 
\be
\Pi_{\ab}=p_{\rm n}\delta_{\ab} 
+ M_{\ab}, 
\en 
where $\alpha$ and $\beta$ stand for 
($x, y, z)$ and   
\bea 
&&\hspace{-1cm} p_{\rm n}=p_{\rm w}+ 
k_BT {\sum}'_{i}(1+n_1\nu'_i )n_i,\\
&&\hspace{-1cm}M_{\ab}= 
\frac{1}{8\pi} \big(\epsilon- {\sum}_i n_i \epsilon_i' 
\big)E^2\delta_{\ab}  -\frac{1}{4\pi} E_\alpha D_\beta ,     
\ena 
The $p_{\rm n}$ is the non-electric solution pressure  
 arising  from the first two terms in $f$ in Eq.(15) 
 \cite{Oka2} with 
 $p_{\rm w}(n_1,T)= \p f_{\rm w}/\p n_1- f_{\rm w}$ 
being   the pure-solvent pressure. The derivatives $\nu_i'$ are related to 
the solute partial volumes (see Eq.(51)). 
The   $M_{\ab}$ is the 
 Maxwell stress tenor of the solution \cite{Landau-e} with 
$D_\beta= \epsilon_{\rm eff}E_\beta$ 
and  $M_{\ab}= M_{\beta\alpha}$, where   
$\epsilon_i'$ are defined in  Eq.(29). 
In its  first term, 
 $\epsilon- {\sum}_i n_i \epsilon_i'$ 
is equal to $ \epsilon_{\rm w}-n_1\epsilon_1'$ 
 to leading order in $n_i$ ($i\ge 2$)  
from $\epsilon$ in    Eq.(5). It is also equal to 
$ \epsilon_{\rm eff}^0- {\sum}_i n_i 
\p\epsilon_{\rm eff}^0/\p n_i$ in terms of $ \epsilon_{\rm eff}^0$ in 
 Eq.(14), so Eq.(39) 
surely tends to  the linear polarization 
limit   $ \epsilon_{\rm eff} \to  \epsilon_{\rm eff}^0$ 
in the Landau-Lifshits book. 
These authors  presented  $M_{\ab}$ for 
one-component dielectric fluids without ions in  applied electric field 
(see Eq.(15.9) in their book \cite{Landau-e}). In this paper, we use it  
for dilute electrolytes containing  dipoles 
with nonlinear polarization.

The reversible (non-dissipative) 
force density  acting on the fluid  
is given by $-{\sum}_\beta \nabla_\beta 
\Pi_{\ab}$. From Eqs.(4), (12),  and (39)  its electric part 
is calculated as      
\bea 
&&\hspace{-1cm}-{\sum}_\beta \nabla_\beta M_{\ab}= 
\rho E_\alpha +{\bi p}_d\cdot\nabla E_\alpha 
\nonumber\\
&&\hspace{-15mm} + 
{\sum}_i\nabla_\alpha (n_i \epsilon_i' E^2 )/8\pi-
{E^2}  \nabla_\alpha \epsilon/8\pi,
\ena 
where $\nabla_\alpha$ and $\nabla_\beta$ stand for  
the components of $\nabla =(\p/\p x, \p/\p y,\p/\p z)$. 
The second  term can also be written as 
$\chi_d \nabla_\alpha E^2/2$ from $\nabla_\beta E_\alpha= 
\nabla_\alpha E_\beta$. 
In ${\sum}_\beta \nabla_\beta 
\Pi_{\ab}$, the pressure gradient  $\nabla_\alpha p_{\rm n}$ is equal to 
 the non-electric part of the combination 
${\sum}_i  n_i \nabla_\alpha \mu_i + S\nabla_\alpha T$ 
   from the Gibbs-Duhem relation. We further confirm that 
${\sum}_\beta \nabla_\beta M_{\ab}$ in Eq.(40) 
coincides with the   electric   part 
 of this combination  from   Eqs.(22)-(25) and (30). 
 Thus,   we obtain  
\be 
{\sum}_\beta \nabla_\beta\Pi_{\ab}= 
{\sum}_i  n_i \nabla_\alpha \mu_i 
+ S\nabla_\alpha T,   
\en 
 including the electric parts. 
The mechanical equililbrium 
 ${\sum}_\beta \nabla_\beta\Pi_{\ab}= 0$ 
is attained for   homogeneous  $\mu_i$ and $T$. 
We note that  Eq.(41)  can be used in 
 the hydrodynamic equations  of nonequilibrium  electrolytes.

  We  can calculate the pressure contribution  from the 
Debye-H\"uckel  free energy $f_{\rm DH}= -
k_BT\kappa^3 /12\pi$ in the form 
$p_{\rm DH}= (1-3 n_1\p \ln \epsilon_{\rm w}/\p n_1)f_{\rm DH}/2$ 
\cite{Oka2}, 
where $\kappa$ is the Debye wave number. 
We can also derive this  $p_{\rm DH}$  
from  the   thermal average 
of $M_{ij} $  using the 
Debye-H\"uckel structure factor  of  
the  charge density   \cite{McQ,Stokes,Debye}. 
We note that   $M_{ij} $  can be used in its nonlinear form  for 
the  thermal charge density  fluctuations  
at small  wave numbers   
in agreement with the Debye-H\"uckel theory.

\subsection{Equilibrium solute densities }
In equilibrium, 
the chemical potentials $\mu_i$ and the temperature $T$ 
are homogeneous constants. If we neglect the wall and 
 image potentials,  Eqs.(23)-(25) yield    
\bea 
&&\hspace{-1cm}
n_2= n_2^0 \exp(-\nu_2 -U+\zeta_2),\\
&&\hspace{-10mm} 
n_3= n_3^0 \exp(-\nu_3 +U+\zeta_3), \\
&&\hspace{-1cm}
n_4= n_4^0 {\cal N}_{\rm d}  \exp(-\nu_4+\zeta_4), 
\ena 
where  $n_i^0$ ($i\ge 2)$ are constants independent of space, 
 $\zeta_i$ are   defined in Eq.(26).   
In Eqs.(42) and (43), $\zeta_i$  
 decrease $n_i$  for  $\epsilon_i^{\rm w}<0$. 
Indeed,   we find   $\exp(\zeta_i)=\exp( -0.05  h^2)$  ($i=2,3$) 
 in   ambient liquid water, where we set 
   ${\hat \gamma}=3.5$ and 
  $\epsilon_2^{\rm w}=\epsilon_3^{\rm w}$ in Eqs.(5) and (6) and 
$\mu_0=e \ell_B/2$ in Eq.(10).   
Despite this factor,   $n_3$ ($n_2)$ should  increase  
 near positively (negatively) charged surfaces,  
 owing  to  the factor $e^U$ ($e^{-U}$). In Eq.(44), 
 the factor ${\cal N}_{\rm d}(>1)$ in Eq.(9) 
comes from $-W$ in Eq.(25) 
 amplifying   $n_4$. 
AO  group  \cite{An1} 
found that the dipole density is multiplied  by 
 ${\cal N}_d$  (without  $\nu_i$ and $\zeta_i$ in their 
 expressions).

 From Eqs.(12) and (44) 
the    polarization  ${\bi p}_d$ becomes 
\be 
{\bi p}_d= \mu_0 n_4^0  e^{-\nu_4+\zeta_4} {\cal G}(h)
 E^{-1}{\bi E}, 
\en 
where ${\cal G}(h)$ is defined by \cite{An1,An2,An4,An3}  
\be 
{\cal G}(h)=\p {\cal N}_d/\p h  ={\cal N}_d {\cal L}(h) =
  \cosh h/h- \sinh h/h^2.
\en  
 Now, 
Eq.(4) gives  the  four-component dPB equation,  
\be 
\nabla\cdot\epsilon{\bi E} 
+ 4\pi\nabla\cdot{\bi p}_d= 4\pi Ze (n_2-n_3)  , 
\en 
where $n_i$ and ${\bi p}_d$ are  given in Eqs.(42)-(45). 
We use   the linear dielectric constant 
  $\epsilon$ in Eq.(5), which  depends on  $n_2$ and $n_3$. 
The decrease of $\epsilon$ 
due to accumulation of cations or anions 
is crucial in solving Eq.(47) \cite{Roij1}, 
though not studied in this paper.   

\subsection{Density change of 
nearly incompressible solvent and 
osmotic stress}

Next, we examine the equilibrium solvent density $n_1$  
in a  nearly incompressible solvent 
in a one-phase state. Its deviation is small if  
 the solvent  isothermal compressibility 
$\kappa_T^{\rm w}= (\p n_1/\p p)_T/n_1$ is much smaller 
than $1/k_BT n_1$. For example,  $k_BT n_1\kappa_T^{\rm w} \cong 
0.06 $  in  ambient liquid water.  We can
 then  expand $\mu_{\rm w}(n_1,T)$ in Eq.(22)    as  
\be 
\delta\mu_{\rm w}= 
\mu_{\rm w}(n_1,T)-\mu_{\rm w}(n_1^0, T)
=  (n_1-n_1^0)/n_1^2\kappa_T^{\rm w} ,\\
\en  
where    $n_1^0$ is  a constant  reference density.
 
For simplicity, let   the electrolyte be in equilibrium with  
a large  pure solvent without applied field,  
for which $n_1^0$ is  the solvent density in the 
reservoir.  Here,  the solvent chemical potential 
is  common in  the two regions as 
\be 
\mu_1(\{ n\},T) = \mu_{\rm w}(n_1^0,T),
\en  
so Eq.(22) yields   the solvent density deviation,  
\bea 
&&\hspace{-16mm}
n_1/n_1^0-1 = -n_1 \kappa_T^{\rm w}k_BT[{\sum}_i' \nu_i' n_i -\zeta_i]
\nonumber\\
&& 
= -{\sum}_i' v_i^*n_i + n_1\kappa_T^{\rm w}{ \epsilon_1'} 
 E^2/8\pi .
\ena  
For  each   $i\ge 2$,  $v_i^*$ is a  volume 
related to the partial volume 
${\bar v}_i$   at infinite dilution \cite{Buff,Oc,Oka1,Oka2} as 
\be 
v_i^*= n_1k_BT\kappa_T \nu_i'= {\bar v}_i- k_BT\kappa_T.  
\en 
The difference ${\bar v}_i -v_i^* = k_BT\kappa_T^{\rm w}$ 
 stems   from the  partial pressure $k_BT n_i$ and is small, which 
 is $0.06/ n_1$ in  ambient liquid water. 
In  Eq.(50),  the first term represents  
 the steric effect \cite{Oka2} and the second term    the 
electrostriction  \cite{Landau-e,Stell,Hakim}. 
Note that   $v_i^*$ can be negative due to the hydration 
(see the last paragraph in this subsection).   
On the other hand, in the  lattice theory of fluid mixtures,  all 
the particles  have a common  volume $v_0$ 
and  the space-filling relation  
 ${\sum}_i v_0 n_i=  1$ is assumed \cite{PG,Iglic,An5}.

In the above situation we also calculate 
the  equilibrium stress component $\Pi_{zz}$  
 in the one-dimensional geometry, where   
 the electric field  ${\bi E}$ changes  
 along the $z$ axis. Here, 
$p_{\rm w}(n_1,T)=  p_0 + n_1\delta\mu_{\rm w}$, 
with $p_0$ being the reservoir pressure. 
From  Eqs.(37)-(39), (48), and (50) we thus find \cite{Is}  
\be 
\Pi_{zz}=p_0 + k_BT {\sum}_i' n_i 
 + (\epsilon_{\rm w}-2\epsilon_{\rm eff})E_z^2/8\pi  . 
\en 
From Eqs.(42)-(44) we confirm that 
 the above $\Pi_{zz}$ is  independent of 
$z$ (or $d\Pi_{zz}/dz=0$) in  accord with Eq.(41).   

We comment on the partial volumes of salts 
in ambient liquid water.   In experiments 
 with the overall charge neutrality, 
the sum ${\bar v}_s= {\bar v}_2+{\bar v}_3$ 
has been measured. From data of  ${\bar v}_s$ \cite{Mil,Craig}, 
$n_1({ v}^*_2+{ v}^*_3)$ was  $-0.20$,  $-0.55$, $0.93$, and $2.0$  
 for LiF, MgSO$_4$, NaCl, and NaI,   
respectively. Then, $n_1(\nu_2'+\nu_3')$ is  
 $-3.4$, $-10$, $15$, and $32$,  respectively, for these salts. 
For small and/or multivalent ions, 
$v_i^*$ and $\nu_i'$ are  negative, while 
 $\nu_i'\gg 1$  for large ions.

\section{Solvent nonlinear polarization}

In this section we consider equilibrium 
three-component electrolytes composed of a solvent, cations, and anions, 
where the solvent exhibits nonlinear 
polarization and  is nearly incompressible.   Indeed, 
AO group   treated such  electrolytes  
\cite{An1,An2,An3,An4}. 
In this case,    the linear dielectric constant $\epsilon_{\rm w}$ 
of the solvent is given by 
\be 
\epsilon_{\rm w}=1+4\pi\mu_0^2n_1/3k_BT.
\en 
Since the linear dielectric constant $\epsilon$ 
of the solution is given in  Eq.(5), 
 the nonlinear dielectric constant 
$\epsilon_{\rm eff}$ of the solution is expressed as 
\be 
\epsilon_{\rm eff}= 
1+ \epsilon-\epsilon_{\rm w}  
+ n_1{\mu_0 }E^{-1} {\cal L}(h).  
\en 
where $\epsilon-\epsilon_{\rm w}\cong 
\epsilon_2^{\rm w}n_2+\epsilon_3^{\rm w}n_3$.  
See  $\epsilon_{\rm eff}$ in Eq.(13) also. 
The electric free energy density is then given by  
\be 
  f_e= (1+ \epsilon-\epsilon_{\rm w}) E^2/8\pi 
+    k_BTn_1( h W' -  W),  
\en  
which tends to $\epsilon E^2/8\pi$ as $h\to 0$ from Eq.(53). 
The increment  $\delta f_e$ is obtained  if we replace 
$\delta\epsilon$ by $\delta(\epsilon-\epsilon_{\rm w})$, 
$\delta n_4$ by $\delta n_1$, and   $n_4$  by $n_1$ in Eq.(19). 
 Then, it follows  
 the solvent chemical potential,  
\be 
\mu_{1} = \mu_{\rm w}(n_1,T) +k_BT    
( \nu'_2n_2+\nu'_3n_3)  -k_BT  W, 
\en 
which should be compared with   $\mu_1$ in Eq.(22).
The expressions for $\mu_2$ and $\mu_3$ are still  given by 
Eqs.(23) and (24). 
Notice that the last term $ -k_BT  W$ in Eq.(56) tends to  
the electrostriction term $-k_BT\zeta_1$ 
  as  $h\to 0$ from  
\be 
k_BTW\cong k_BTh^2/6=  (\p \epsilon_{\rm w}/\p  n_1) E^2/8\pi
~~ (h\ll 1).
\en

We should  consider the deviation of $n_1$. 
Let us    determine the reference solvent density $n_1^0$ 
from the osmotic condition (49). 
As in  Eq.(50), we find   
\be 
n_1/n_1^0-1 = - v_2^*n_2-v_3^*n_3
 +(k_BT n_1\kappa_T^{\rm w})W.  
\en 
The  last term  is  small 
from $k_BT n_1\kappa_T^{\rm w}\ll 1$ 
and $W\cong h$ for $h\gg 1$ (see the first paragraph 
in Sec.IIE).  
That is,   the polarization-induced increase in $n_1$ 
is negligible for small solvent compressibility, while that in $n_4$ can be significant 
as found in Sec.II. 
Thus, regarding $n_1$ as a constant,  we    
now set up the three-component  dPB equation as     
\be 
\nabla\cdot\big[1+ \epsilon-\epsilon_{\rm w}  
+ n_1{\mu_0 }{E}^{-1} {\cal L}(h)\big]{\bi E}
= 4\pi Ze (n_2-n_3)  ,
\en 
where  $n_4^0e^{-\nu_4+\zeta_4}$ is replaced by $n_1$ 
and ${\cal G}={\cal N}_d{\cal L}$   by   ${\cal L}$ in Eq.(45). 
Here, we  neglect  the steric effect 
due to the first two terms in the right hand side of Eq.(58).

 In  their three-component dPB equation, 
 AO group  used ${\cal G}$ instead  of $\cal L$,  
where the  polarization  contribution 
is larger  than in Eq.(59) by ${\cal N}_d$. 
See Eq.(5) in Ref.\cite{An1}  and   Eq.(27) in Ref.\cite{An3}.   
They further  accounted for the steric effect 
 dividing  the polarization contribution 
and the charge density 
by ${\cal D}= 
(1-\phi_0){\cal N}_d + \phi_0 \cosh(U)$, 
where $\phi_0$ is the bulk ion  volume fraction.   
See Eq.(9) in Ref.\cite{An1} and Eq.(13) in Ref.\cite{An4}.   
However, as  indicated by  Eq.(58), 
we should  divide them  by \cite{An5,Iglic} 
\be 
{\cal D}_0= 1-\phi_0 + \phi_0 \cosh(U), 
\en   
where  all the particles are assumed to have  the same volume 
(see below Eq.(51)). The steric effect is 
crucial in high-potential regions with   $\phi_0 \cosh(U)\gs 1$.

\section{Ion association and dissociation}

In this section, we examine the equilibrium electric-field effect 
for  Bjerrum dipoles composed of 
monovalent ions \cite{Bje,Fuoss,Roij,Ma,Dill,Vegt}.  
For simplicity,   we assume that they have a fixed  
dipole length $\ell_d$ with a constant  dipole moment,  
\be 
\mu_0 = e \ell_d,  
\en 
though $\ell_d$ should thermally 
fluctuate  due to rather weak bonding. 
 In  the original theories of ion pairing \cite{Ma,Bje,Fuoss}, 
 $\ell_d$ is of order $\ell_B$.  We then use the Onsager theory of the 
orientation  entropy in Appendix A.
For relatively high  salt concentrations ($\gs 1$M$/\ell$), 
clustering of ions becomes conspicuous in simulations 
\cite{Vegt,Ma,Deg,Hassan}.

On   symmetric multivalent  ion pairs (A$^{+Z}$B$^{-Z}$), 
we make comments here and in the last paragraph 
of this section. The  dipole 
moment of such ion pairs  is   of order  $Z^3 e\ell_B$ 
if there separation  length  is  
of order $Z^2\ell_B$. 
   However, this picture is not justified   
because a number of water molecules are 
nonlinearly oriented between them. 
Nevertheless, the   electric field effect 
should be enhanced  for multivalent ion pairs. 
 It is worth  noting that the chemical reaction of 
 MgSO$_4$ is the main origin of   
low-frequency sound  attenuation in sea water 
  \cite{Yeager1,Sim,Eigen},    
where  the dissociation time 
  is of order  $10^{-6}$sec and is very long.

\subsection{Chemical equilibrium }

In chemical thermodynamics \cite{Landau-s,Kirk}, 
the condition of chemical equilibrium is given by 
$ \mu_2+\mu_3= \mu_4$, where dissociation and association 
balance on the average.  This condition 
is rewritten as 
\be 
 \mu_2+\mu_3- \mu_4 =k_BT \ln [Kn_2n_3/n_1n_4]=0, 
\en 
where $K$ is the equilibrium  association constant 
in the dimensionless form. 
 From Eqs.(23)-(25) we have  
\be 
\ln K
=  \ln[n_1(\lambda_2\lambda_3/\lambda_4)^3] 
+ \nu_r  -W- \epsilon_r'  E^2/(8\pi k_BT) , 
\en  
where  $\nu_r$ and $\epsilon_r' $ are defined by     
\bea 
&&\hspace{-1cm}\nu_r = \nu_2+\nu_3-\nu_4, \\
&&\hspace{-1cm} \epsilon_r'
=\epsilon_{2}'+\epsilon_3'-\epsilon_4'  
\cong \epsilon_{2}^{\rm w}+\epsilon_3^{\rm w}-\epsilon_4^{\rm w} . 
\ena  
In experiments,  $K$  is determined in   equilibrium by  
\be 
K= n_1 n_4/n_2 n_3= c_4/c_2 c_3,
\en 
where $c_i=n_i/n_1$ $(i\ge 2$) 
are  the solute concentrations. 
Typically,    $K=10^2$ 
for $Z=1$  and $K=10^4$ for $Z=2$ without applied field 
\cite{Ma,Rie}.

In $f$ in Eq.(15),  $\nu_4$ is the parameter for  the Bjerrum dipoles. 
If there is no  applied field, its non-ideal part $\nu_4-\nu_4^0 $ 
is determined by    the  Coulombic 
attraction  between 
an associating  ion pair   under the 
influence of  the solvent. 
For small  1:1 ion pairs,     the mutual  potential 
is given by $-(k_BT \ell_B)/r $ 
if the   separation distance $r$  is shorter than 
$\ell_B$ \cite{Bje,Fuoss,Ma,Dill,Vegt,Roij}. 
  Bjerrum \cite{Bje} 
and Fuoss \cite{Fuoss}  estimated  $K$ as 
\be 
K_{\rm B} \sim a^3  \int_a^{\ell_B/2} \hspace{-1mm}  dr r^2
 e^{\ell_B/r}, ~~K_{\rm F}\sim e^{\ell_B/a},  
\en 
respectively.  Here, $a$ is  the closest center-to-center 
distance of an  ion pair.  
In Bjerrum's theory,  the bounded pairs are those 
with $a<r<\ell_B/2$.  If we use  their 
estimations, the association contribution to 
$\nu_4-\nu_4^0$ is given by  $-\ln K_{\rm B}$ or 
$-\ln K_{\rm F}$  in the absence of applied  field 
from Eqs.(63) and (64).

\subsection{Field-dependent association constant}

In applied electric field ${\bi E}$, 
the ion  bonding  tends to be partially broken  
when the dipole direction ${\bi n}_d$ 
 is  parallel to   $\bi E$, while the bonding 
is stronger  for  ${\bi n}_d\cdot \bi E<0$. 
The Wien effect for weak electrolytes 
\cite{Wien,Eck,Kaiser,Korea} 
indicates that  the  dissociation constant $K_d$ 
should  increase   with increasing  $E= |\bi E|$,  
where  $K_d$  is the inverse of the association constant  $K$.
For 1:1 salts,  
Onsager \cite{Onsager49} calculated the $E$-dependence 
of  $K_d$   in the form,  
\be 
K_d(E)/K_d(0) = F_{\rm O}(b)= I_1(\sqrt{8b})/\sqrt{2b}.
\en 
where $I_1(x)=\pi^{-1}\int_0^\pi d\theta \cos\theta \exp({x\cos\theta}) 
$ is the modified Bessel function of the first kind and $b$ is given by 
\be 
b=\ell_B eE/2k_BT=(\ell_B/2\ell_d)h, 
\en 
with $\ell_d$ being  defined in Eq.(61). 
Here,    $F_{\rm O} (b) \cong 1+ b$ for $b\ll 1$  and   
  $F_{\rm O} (b) 
\cong \sqrt{2/\pi} (8b)^{-3/4}
\exp(\sqrt{8b})$ for $b\gg 1$. 
The characteristic field at $b=1$ is 
 $E= 2k_BT /e\ell_B= 0.08$V$/$nm. 
In particular,   $b=h$ for 
 $\ell_d= \ell_B/2$.  
Onsager derived Eq.(68) kinetically from the equations for
Brownian motion in the combined Coulomb and external fields. 
However, $K_d(=1/K)$ is determined by Eq.(62)  in equilibrium, 
so it  can   be calculated 
by  a static theory. 

To understand the Onsager formula (68), let us assume that  ${\bi n}_d$ 
is parallel to   ${\bi E}$.    Then, the sum of 
  the Coulombic   potential   
and  the field potential is written as    
\be 
\psi_B(r)= -k_BT \ell_B /r -eEr.
\en 
which   assumes a maximum  $\psi_B^{\rm max}$ 
at $r= \ell_B /\sqrt{2b} $ with 
\be 
\psi_B^{\rm max}/k_BT= - \sqrt{8b}.  
\en  	
Thus, we find   $\sqrt{8b}$ in  Eq.(68). However, 
in the above  argument (and in  Osager's  calculation also), 
the   thermal random rotations of the dipoles 
 are neglected, which should be important 
particularly   for  weak applied  field.

 In this paper, we treat the  field-induced dissociation  
 as a consequence of 
the field-induced   potential change,  as argued 
in the above paragraph.  That is, we assume that  
 $\nu_4$ in the chemical potential $\mu_4$ 
   is  field-dependent as 
\be 
\nu_4(E)=\nu_4(0) + \ln F_{\rm O}(b) ,
\en 
where we use Onsager's   $F_O$. 
Then,  $\mu_4$ in  Eq.(25)  becomes   
\be 
\mu_4(E)= \mu_4(0)- k_BT \ln ({\cal N}_d/F_{\rm O}) - \epsilon_4'E^2/8\pi.
\en 
Thus, in the presence of electric field, $\mu_4/k_BT$ 
decreases by $W= \ln {\cal N}_d$ 
due to the dipole alignment but increases 
by $ \ln F_{\rm O}$ due to the potential change. 
Here, $W\propto h^2$ and $\ln F_O\propto h$ 
for small $h$.    
In Fig.1(a), we plot $\ln ({\cal N}_d/F_{\rm O})$ 
 for $b=h$ using Onsager's  $F_O$, where 
$F_{\rm O}$  is larger (smaller) than ${\cal N}_d$  
for $h\ls 7$ (for $h\gs 7$). 

For 1:1 salts, Eqs.(63)-(65) 
and (72) yield   the field-dependent 
association constant $K(E)$   in the form,  
\be 
K(E)/K(0)  =\exp( A_r h^2)
{\cal N}_d /F_{\rm O}(b).
\en 
If we use $\hat\gamma$ in Eq.(6) and set $\epsilon_4'=0$, 
 $A_r$  is written as 
\be 
A_r= -(\epsilon_2^{\rm w}+\epsilon_3^{\rm w})  k_BT/8\pi \mu_0^2
=  {\hat\gamma}/(4\pi \ell_B\ell_d^2n_1). 
\en 
In Fig.1(a),  $K(E)/K(0)$ is plotted 
for   $b=h$  and  $A_r= 0.10$,  
 where the factor  $\exp( A_r h^2)$ is important for $h\gs 3$. 
Thus, $K(E)$ decreases 
due to Onsager's  dissociation mechanism 
 for relatively small $h (\ls 3)$, 
but it increases due to the dipole accumulation  
and the ion-induced dielectric degradation 
 for larger $h$.

\subsection{Electric double layer  with Bjerrum dipoles }

\begin{figure}[t]
\includegraphics[scale=0.47]{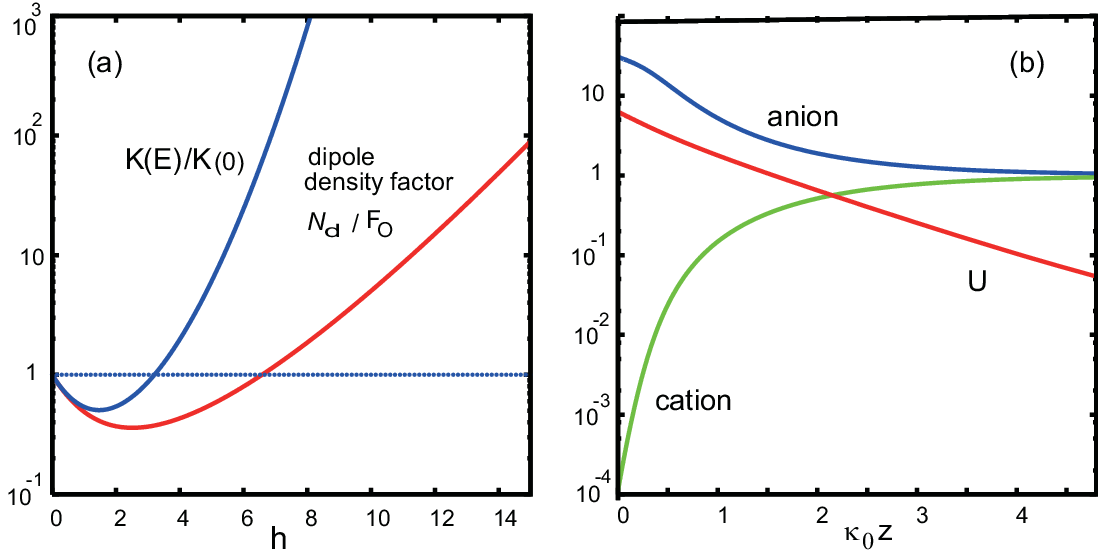}
\caption{\protect  (a) Field-dependence of the  association 
constant $K(E)/K(0)$ in Eq.(74)  
and  the  dipole density factor 
$c_4/c_d^0= {\cal N}_d/F_O$ in Eq.(79)  
as functions of  the normalized field strength 
$h=\mu_0 E/k_BT$ in Eq.(10) 
for 1:1 salts on a semi-logarithmic scale. 
In $\mu_4$ in Eq.(73) $\ln({\cal N}_d/F_O)$ 
also appears.
(b) Near-wall profiles of 
$U(z)$ (electric potential),  $c_2(z)/c_0$ (cation), and  $c_3(z)/c_0$ (anion) 
vs $\kappa_0z$  on a semi-logarithmic 
scale, where the bulk concentrations are 
$c_0= c_d^0=10^{-2} $ and $\kappa_0$ is the bulk Debye wave number. 
}
\end{figure}

We consider an electric double layer 
with a charged surface at $z=0$ 
in the semi-infinite limit ($0<z<\infty$). 
 For large  $z$, 
  $c_2(z) =n_2(z)/n_1 $ and $c_3(z)= n_3(z)/n_1$ tend to  
 $  c_0$, while  $c_4(z)= n_4(z)/n_1$ tends to $ c_d^0$. 
From Eq.(66) the bulk  concentrations $c_0$ and $c_d^0$    satisfy 
\be 
c_d^0 =K(0) c_0^2 .
\en 
Neglecting the steric effect we write 
the solute concentration  profiles  $c_i(z) = n_i(z)/n_1$ as  
\bea 
&&\hspace{-1cm} 
 c_2(z)= c_0 \exp(-U-A_2h^2), \\
 &&\hspace{-1cm} 
 c_3(z)= c_0 \exp(U-A_3h^2),\\
&&\hspace{-1cm}  c_4(z) =c_d^0 
 {\cal N}_d/F_{\rm O}, 
\ena 
where 
 $A_i= -\epsilon_i^{\rm w}  k_BT/8\pi \mu_0^2$ ($i=2,3$) with 
$A_2+A_3=A_r$ for  $\epsilon_4'=0$. 
The  $U(z)$ and $h(z)$  tend  to 0 as $z\to\infty$. 
See  Fig.1(a) for  $c_4/c_d^0= {\cal N}_d/F_O$ vs  $h$.

 In Fig.1(b), we plot $U(z)$ 
$c_2(z)/c_0$, and  $c_3(z)/c_0$ 
vs $\kappa_0 z$  for a 1:1 salt, where 
 $c_0= c_d^0=10^{-2}$, $K(0)= 10^2$,  
and $A_2=A_3=0.05$ with  $\kappa_0$ 
being  the bulk  Debye wave number. 
In this example,  $U$ exhibits a nearly exponential 
decay. The   parameter values are common to those in (a).

Finally, we examine  the dielectric constant $\epsilon_{\rm eff}$ 
 in the bulk region for $Z:Z$ salts. From Eq.(14) it is written as 
\be
\epsilon_{\rm eff}/\epsilon_{\rm w}\cong 
1- 2{\hat\gamma}c_0+ A_d  c_d^0, 
\en  
where we use  Eq.(6) with $\epsilon_4^{\rm w}=0$ in the second term. 
From Eqs.(14) and (57)  the coefficient $A_d$ is given by  
\be 
A_d =4\pi Z^2\ell_B\ell_d^2n_1/3 .  
\en 
For $Z=1$ and $\ell_d=\ell_B/2$,   $A_d$ is about $12$. 
For multivalent ion pairs, $A_d$ is larger (see below Eq.(60)).  
Thus,  the third dipolar term in Eq.(80)  can exceed 1 
for $c_d^0> A_d^{-1}$ or for $c_0> [A_d K(0)]^{-1/2}$, 
where the lower bound of $c_0$ is 0.03 for $Z=1$ 
and $1.5 (\ell_B/2\ell_d)\times 10^{-3}$ for $Z=2$.  
Here, we set $K(0)=10^2$ 
for $Z=1$  and $K(0)=10^4$ for $Z=2$ \cite{Ma,Rie}. 
If $c_0$ much exceeds the lower bound, 
we have $\epsilon_{\rm eff}\propto c_0^2$ 
and the  Debye wave number becomes  proportional to 
$[c_0/\epsilon_{\rm eff}]^{1/2}
\propto c_0^{-1/2}$, as predicted  by Zwanikken  and Roij \cite{Roij}.
We also note that   the second term in Eq.(80) serves to decrease 
$\epsilon_{\rm eff}$,  which can also be important 
with increasing the $c_0$.

\section{Summary and remarks} 
We have studied the effects 
of  electric field in  dilute electrolytes, 
which contain   dipoles exhibiting  nonlinear polarization.  
 Our main results are summarized below.

 In Sec.II, we have discussed 
thermodynamics of electrolytes 
in inhomogeneous electric field, 
where the dipolar component is a dilute solute.  
 The electric  free energy density $f_e$ has been 
presented in Eq.(17) using Onsager's theory \cite{Onsager49}. 
We have then defined the field-dependent  
 chemical potentials $\mu_i$ in Eqs.(22)-(25) 
and the entropy density $S$ in Eq.(28). 
We have calculated the equilibrium density  profiles 
of  the solutes and  the solvent in  Eqs.(42)-(45) and (49). 
 In Sec.III, we have considered  three-component 
electrolytes (solvent+ strong acid), where the solvent 
polarization can be  nonlinear. 
In this case, if the solvent is nearly incompressible, 
 the solvent density increase 
due to nonlinear polarization is negligible, 
which is not in accord with the theory 
by the Andelman-Orland group. 
In Sec.IV, we have examined Bjerrum dipoles 
created by association and 
dissociation of 1:1 salts, which 
  have a large dipole moment. We have calculated  
the density profiles in Eqs.(71)-(73) and 
the field-dependent association constant $K(E)$ 
in Eq.(68).

We make some remarks below. 
(i) The Debye-H\"uckel free energy 
density   and 
 the short-range ion-ion interaction terms 
are needed for not very 
dilute electrolytes \cite{McQ,Stokes,Oka2}.    
 We should further include the steric effect in our theory 
to  account for the accumulation of 
ions and dipoles near charged surfaces \cite{Iglic,An5,Oka2}. 
We have also neglected the short-range solute-wall interactions 
and the image potentials. 
(ii) The decrement of the dielectric constant $\epsilon$ 
due to ions in Eq.(5) should greatly affect 
the ion densities near charged surfaces \cite{Roij1}, 
so its influence should be examined in detail. 
(iii)  The Onsager theory of  field-induced dissociation \cite{Onsager34} 
has not attracted enough attention, so 
his formula (68) should be checked  
in detail  (see our remark  below Eq.(71)). 
(iv) We should study the effect 
of electric field on electrolytes with 
  multivalent salts such as MgSO$_4$ and 
 MgCl$_2$. (v)  Recently, the screening length has been 
shown to be  longer than the Debye 
length $\kappa^{-1}$ in ionic liquids \cite{Is1} and concentrated 
electrolytes ($>1$M$/\ell$ for NaCl) \cite{Perkin1}.  
At present, we cannot decide whether or not the last paragraph 
in Sec.IV is related to this  issue.   
(vi) Phase separation in  electrolytes 
with  mixture solvents are also of great interest 
\cite{Onukipolar,Tsori,elec,anta}, 
where electric field appears across two-phase 
interfaces and is particularly strong 
in the presence of hydrophobic and hydrophilic 
(antagonistic) ions \cite{anta}. 
The role of Bjerrum dipoles 
remains unknown in mixture solvents.

  \noindent 
{\bf Data availability}: The data that supports the findings of
this study are available within the article.

\vspace{2mm}
\noindent{\bf Appendix A: Polarization free energy 
}\\
\setcounter{equation}{0}
\renewcommand{\theequation}{A\arabic{equation}}
In this appendix, we  consider  four-component electrolytes, where  
 the solvent polarization ${\bi p}_{\rm w}$ 
and the  orientational distribution  $P( {\bi n}_d, {\bi r})$ 
for    the dipolar component are  unknown variables to be determined 
below. Here,  the dipole polarization ${\bi p}_d$ 
is given by Eqs.(1) and (7) in terms of $P( {\bi n}_d, {\bi r})$ 
with  $ \int  {d\Omega} P=4\pi$.

We  minimize ${\cal F}$ at fixed surface charges.   
The electric free energy density is given by   
\be 
f_e= \frac{1}{8\pi}E^2 + \frac{1}{2\chi} p_{\rm w}^2 
- T n_4 s_{\rm d}. 
\en 
where  $\chi= (\epsilon-1)/4\pi$ and $p_{\rm w} =|{\bi p}_{\rm w}|$. 
If we neglect  the short-range interactions  among the 
dipolar particles,    $s_d$ 
is  the orientation entropy per 
 dipolar particle \cite{Onsager49},  
 \be
s_{\rm d}=-\frac{1}{4\pi}    k_B 
 \int\hspace{-1mm} {d\Omega} P\ln  P,  
\en 
where   $s_{\rm d}=0$ for 
the isotropic distribution $P=1$. 

We then superimpose  small variations on 
the physical quantities, where ${\bi p}_{\rm w}$ 
 is changed by $\delta{\bi p}_{\rm w}$ and 
$P$ by $\delta P$. Since 
 $\delta {\bi\mu}_d=\mu_0 \int d\Omega {\bi n}_d \delta P/4\pi$, we obtain    
\bea 
&&\hspace{-6mm}
\delta f_e=  \frac{{\bi E}\cdot \delta {\bi D}}{4\pi} 
- { p}_{\rm w}^2  \frac{ \delta\chi}{2\chi^2}
+  \Big(\frac{{\bi p}_{\rm w}}{\chi} -{\bi E}\Big)
\cdot\delta{\bi p}_{\rm w}  -   s_{\rm d} \delta(Tn_4)  \nonumber\\
&&\hspace{-4mm}-{\bi E}\cdot{\bi\mu}_d\delta n_4
+   \frac{n_4}{4\pi}\hspace{-1mm} \int \hspace{-1mm} 
{d\Omega}  \Big(k_BT\ln P-\mu_0{\bi E}\cdot {\bi n}_d\Big)\delta P .
\ena 
Thus, to minimize $f_e$, we obtain    
${\bi p}_{\rm w}= \chi{\bi E}$ and 
  $P= \exp(\mu_0{\bi E}\cdot {\bi n}_d/k_BT)/{\cal N}_d$, 
under which the third and last terms  vanish and the 
second term becomes $-E^2\delta\epsilon/8\pi$ in Eq.(A3). 
Then, we have  Eqs.(2) and (8).   
For  this equilibrium $P$  we find  
\be 
Ts_{\rm d} 
= -{\bi E}\cdot{\bi \mu}_d + k_BT \ln {\cal N}_{\rm d}
= k_BT ( W-hW'),     
\en 
which leads to  Eq.(17). 
We can see $s_{\rm d}\le 0$ and $ds_{\rm d}/dh= -k_B h 
(d{\cal L}(h)/dh) \le 0$, 
In the three-component case in Sec.III, 
$f_e$ is given by Eq.(55).

\end{document}